\documentclass[12pt]{article}
\usepackage{amssymb}
\usepackage{epsfig}
\usepackage{sint,cite}
\usepackage[usenames]{color}
\definecolor{darkgreen}{rgb}{0, 0.6, 0}

\newcommand\be{\begin{equation}}
\newcommand\ee{\end{equation}}
\newcommand\bea{\begin{eqnarray}}
\newcommand\eea{\end{eqnarray}}

\newcommand{\rmO}{{\rm O}}

\def\DH{\Delta H}

\def\Iam1{\langle\exp(-\DH)\rangle}

\newcommand{\eq}[1]{eq.~(\ref{#1})}
\newcommand{\fig}[1]{Fig.~\ref{#1}}
\newcommand{\tab}[1]{Table~\ref{#1}}
\newcommand{\sect}[1]{Section~\ref{#1}}

\def\fm{\,{\rm fm}}

\newcommand{\ev}[1]{\langle #1 \rangle}
\def\rmd{{\rm d}}
\def\ff{{[\phi_1]_{\rm F}}}

\def\ma[#1,#2,#3,#4]  {{\left( \matrix{ #1  & #2 \cr
                                        #3  & #4 \cr } \right)}}

\begin{document}

\thispagestyle{empty}
\title{{\normalsize\vskip -50pt
\mbox{} \hfill HU-EP-04/27 \\
\mbox{} \hfill SFB/CPP-04-15 \\}\vskip 25pt
Impact of large cutoff--effects on algorithms\\
for improved Wilson fermions\\
\vskip 25pt
}\author{
\centerline{
            \epsfxsize=2.5 true cm
            \epsfbox{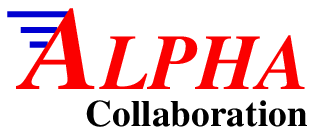}}\\
Michele Della Morte, Roland Hoffmann,\\
Francesco Knechtli and Ulli Wolff\\[1cm]
Institut f\"ur Physik, Humboldt Universit\"at,\\
Newtonstr. 15, 12489 Berlin, Germany\\[1.0cm]
}

\date{July 7, 2004}

\maketitle

\begin{abstract}
  As a feasibility study for a scaling test
  we investigate the behavior of algorithms for
  dynamical fermions in the $N_f\!=\!2$ Schr\"odinger
  functional at an intermediate volume of $1\fm^4$.
  Simulations were performed using HMC with two
  pseudo--fermions and PHMC at lattice spacings of
  approximately $0.1$ and $0.07\fm$. We show that some algorithmic
  problems
  %arising at these parameter values
  are due to large cutoff--effects in the spectrum of
  the improved Wilson--Dirac operator and disappear at
  the smaller lattice spacing. The problems discussed here
  are not expected to be specific to the Schr\"odinger
  functional.
\end{abstract}

\thispagestyle{empty}

\newpage

\section{Introduction \label{intro}}

\subsection{Motivation}

For improved Wilson fermions it has long been established that
in the quenched approximation cutoff--effects at a lattice
spacing of $a\!\simeq\!0.1\fm$ are tolerable
and a continuum extrapolation can be started there.
Recently more and more evidence has been accumulated that for
dynamical improved Wilson fermions in a similar physical condition
the cutoff--effects are much larger than expected. As an extreme
example, for three flavors the existence of a phase transition in the
$\beta$--$\kappa$--plane has been numerically conjectured and is 
interpreted as a lattice artifact \cite{Aoki:2001xq}.
A summary of large scaling violations in the two--flavor--theory
is given in ref.~\cite{Sommer:2003ne}.

In order to quantify those we are preparing a scaling test similar to
what was done for the quenched case in ref.~\cite{Heitger:1999dw}.
This will also serve as a benchmark for new actions.
In the course of the scaling study we plan to
calculate the axial current normalization constant $Z_{\rm A}(g_0^2)$ and
the axial current improvement constant $c_{\rm A}(g_0^2)$ using the methods
described in refs.~\cite{Hoffmann:2003mm} and \cite{Durr:2003nc},
respectively.

On our coarser lattices we encountered algorithmic difficulties in
both the molecular dynamics integration of the Hybrid Monte Carlo (HMC)
and the efficient simulation of the canonical ensemble.
We thus found it advantageous to deviate from importance sampling.
Here we will discuss these
problems and their link to cutoff--effects at the infrared end of the
spectrum of the Dirac operator.

This paper is organized as follows: In the remainder of \sect{intro}
we will briefly describe our setup and give a summary of the parameters
of the simulations, which we will quote in the following sections.
In \sect{sect_two} we describe in more detail the problems encountered
at large coupling and also discuss methods to address these. In this context
we study the behavior of the 
Polynomial Hybrid Monte Carlo (PHMC) algorithm
\cite{deForcrand:1996ck,Frezzotti:1997ym}
in this situation and find it a very useful tool for a detailed
investigation of the properties of the small eigenvalues of the
Dirac operator.

\sect{sect_three} is devoted to a comparison of the spectrum
with the quenched case at matched physical parameters and in
\sect{sect_four} we present results from an exploratory run at a smaller
lattice spacing. We close with a brief summary of our findings.

\subsection{Setup}

All our simulations were performed in the Schr\"odinger functional (SF)
setup \cite{Luscher:1992an,Sint:1993un}.
We use non--perturbatively improved Wilson fermions
\cite{Luscher:1996ug,Luscher:1996sc,Jansen:1998mx,Yamada:2004ja} and
the plaquette gauge action. 
For $N_f\!=\!2$ the clover coefficient $c_{\rm SW}$ at $\beta\!=\!5.2$ has been set
to the value suggested in \cite{Jansen:1998mx} and recently confirmed
by the JLQCD Collaboration \cite{Yamada:2004ja}.
For the additional boundary--improvement
coefficients needed in the SF we used the perturbative values for
$c_t$ (2--loop) \cite{Bode:1999sm} and $\tilde c_t$ (1--loop) \cite{Sint:1997jx}.
The axial current improvement
constant $c_{\rm A}$ is also set to its 1--loop value \cite{Luscher:1996vw}.
The first algorithm used
is the HMC with two pseudo--fermion fields as proposed in
ref.~\cite{Hasenbusch:2001ne}. We want to note that
the physical situation here is quite different from the one where this
algorithm was previously applied and its performance tested by the
ALPHA Collaboration
\cite{DellaMorte:2003jj}. In our planned scaling study we are interested in
intermediate size physical volumes and lattice spacings between
$0.1$ and $0.05\fm$.
As mentioned above the second algorithm we employed
is the PHMC, which we will discuss in some detail
in \sect{sect_two}. Apart from global sums all our calculations are
carried out in single--precision arithmetics. 

\subsection{Simulation parameters \label{simpar}}

\noindent
In \tab{t_simpar} we list the lattice sizes and bare parameters of our simulations.
In all cases we have $T\!=\!9/4\;L$ for the time extension $T$.
The bare quark mass $m$ is defined in the appendix of
ref.~\cite{Capitani:1998mq}.
In the algorithm column '$\rm H_2$' refers to HMC with two pseudo--fermion fields and
'$\rm P_n$' stands for PHMC with a polynomial of degree $n$. The trajectory length is
always equal to one and the molecular dynamics integration step--size is denoted by
$\delta\tau$. For each simulation we ran 16 independent replica to gain more statistics.
Concerning the SF parameters we work with zero background field
and periodic spatial boundary conditions ($\theta\!=\!0$).

\begin{table}[!h]
\centering
\small
\begin{tabular}{r|ccccllccc}
    \hline
    run & $L/a$ &  $\beta$ & $\kappa$ & $c_{\rm SW}$ & $\ \ Lm$ & algo. & 
    $N_{\rm traj}$    &   $\delta\tau$ & acc.\\
    \hline
\rm I   & 8 & 5.2 & 0.13550 & 2.017   & 0.205(10)  & $\, \rm H_2$ & $16\!\cdot\!500$ &    1/16 & 91\% \\% m0
\rm II  & 8 & 5.2 & 0.13515 & 2.017   & 0.307(9) & $\, \rm H_2$ & $16\!\cdot\!520$ &  1/25 & 97\% \\% m4
\rm III & 8 & 5.2 & 0.13515 & 2.017   & 0.314(8) & $\, \rm P_{140}$  & $16\!\cdot\!500$ & 1/26 & 87\% \\\hline% l08t18/m0
\rm IV  & 8 & 5.2 & 0.13550 & 2.017   & 0.195(7) & $\, \rm P_{140}$  & $16\!\cdot\!400$ & 1/25 & 85\% \\% l08t18/m1
\rm V   & 8 & 6.0 & 0.13421 & 1.769   & 0.193(3) & \multicolumn{4}{c}{--- quenched ---} \\\hline 
\rm VI  &12 & 5.5 & 0.13606 & 1.751   & 0.287(3)& $\, \rm H_{2}$ & $16\!\cdot\!240$ & 1/20 & 91\% \\% l12t27/m1
\rm VII &12 & 6.26& 0.13495 & 1.583   & 0.295(3)& \multicolumn{4}{c}{--- quenched ---} \\     \hline
\end{tabular}
\caption{Summary of simulation parameters.}
\label{t_simpar}
\end{table}

\section{Sampling problems on coarse lattices\label{sect_two}}

\subsection{Instabilities in the molecular dynamics integration}
\label{instab}

Algorithms making use of molecular dynamics (MD) require a numerical integration
of the equations of motion. Along a trajectory the Hamiltonian is then
only conserved up to powers of the step--size $\delta\tau$ employed
in the integration. Apart from these small deviations, under certain conditions
the currently used integration schemes can become unstable and produce very
large Hamiltonian violations $\Delta H$. For a more detailed discussion see
ref.~\cite{Joo:2000dh}, where a connection between these instabilities and
large driving forces in the MD is proposed in analogy to a harmonic
oscillator model.
In this model the integrator becomes unstable when the product of the force
and the integration step--size exceeds a certain value.

The reversibility of the numerical integration is needed to prove detailed balance
for these algorithms, which in turn implies that $\ev{e^{-\Delta H}}\!=\!1$.
Here one should note that the average is taken over \emph{all} proposed
configurations (see ref.~\cite{DellaMorte:2003jj}). Therefore this quantity is
also sensitive to those, which were rejected in the
Metropolis step following the MD integration, i.e.
trajectories resulting in a large value of $\Delta H$.
In a histogram of $e^{-\Delta H}$ these contribute to bins close to zero
while the distribution is peaked around one.
They can also lead to an unusual autocorrelation of
this quantity, making the Monte Carlo error estimate
difficult.\footnote{All our data analysis is done using an explicit
integration of the autocorrelation function as detailed in
ref.~\cite{Wolff:2003sm}.
This method also provides an estimate of
the error of $\tau_{\rm int}$.}
In particular this applies also to the
integrated autocorrelation time of $e^{-\Delta H}$ itself. 
This is due to the long periods of rejection in the Metropolis step,
which sometimes follow large $\Delta H$ values.

\begin{figure}[t]
\begin{center}
\includegraphics[angle=0,width=13cm]{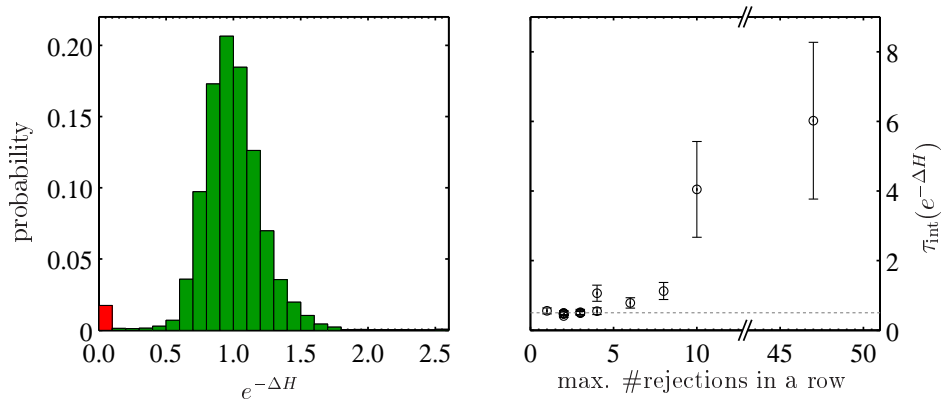}
\end{center}
\vspace{-0.5cm}
\caption{$e^{-\Delta H}$ from run I. Left plot: histogrammed distribution.
Right plot: our estimates of $\tau_{\rm int}$ in units
of MD time separately for the 16 replica. In our normalization
$\tau_{\rm int}=0.5$ means no autocorrelation (dotted line).}
\label{tau}
\end{figure}

\fig{tau} shows a histogram of $e^{-\Delta H}$ and also its integrated
autocorrelation time from one of our simulations.
In this data set there are several series of large $\Delta H$ values, during
which the proposed configurations were rejected.
In the distribution of $e^{-\Delta H}$ these lead to an additional peak close
to zero.
One also sees from the right--hand plot that $e^{-\Delta H}$ is noticeably
autocorrelated only when a large number of proposals were rejected in a row.
As argued above in these cases the error of $\tau_{\rm int}$ could be
underestimated. These two effects might cause some concerns when using
$\langle e^{-\Delta H}\rangle-1$ as an indicator for the absence of
reversibility violations \cite{DellaMorte:2003jj}.

Spikes in $\Delta H$ have been observed by
several collaborations using (improved) Wilson fermions in various setups
(e.g.~different gauge actions and volumes)
at relatively large lattice spacings
\cite{Joo:2000dh,Jansen:1998mx,Namekawa:2002pv,Allton:2004qq}.
There these spikes have been traced back to large values of the driving force in the
MD evolution and also their dependence on the quark mass has been investigated.

\begin{figure}[t]
\begin{center}
\includegraphics[angle=0,width=11cm]{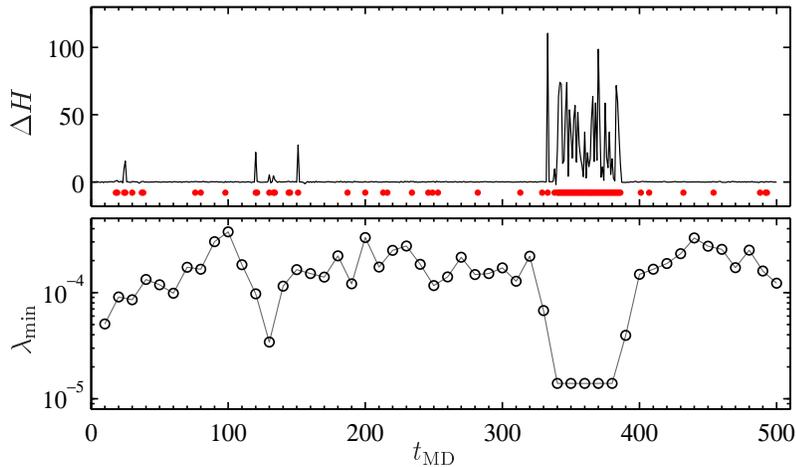}
\end{center}
\vspace{-0.5cm}
\caption{Monte Carlo history for one replicum of run I with a long
period of rejection. Configurations where the new proposal was
rejected are marked by a dot. At $t_{\rm MD}\!\!=\!340$
the algorithm gets stuck with a configuration carrying an exceptionally low
smallest eigenvalue $\lambda_{\rm min}$ of $\hat Q^2$.}
\label{lmin}
\end{figure}

Here we want to clarify a point, which is essentially implied by the
previous observations \cite{Namekawa:2002pv,Frezzotti:2002iv}, namely
the strong correlation
between spikes in $\Delta H$ and small eigenvalues of the Dirac
operator.\footnote{Here and in the following we will always refer to the eigenvalues
of the square of the Hermitian even--odd preconditioned Dirac
operator $\hat Q^2$ in the Schr\"odinger functional. For its precise definition
see ref.~\cite{DellaMorte:2003jj}.}
In this way we hope to be able to separate physical
effects from cutoff--effects, i.e. the occurrence of unphysically
small eigenvalues. In \fig{lmin} we clearly see a long period of rejection
(corresponding to the rightmost data point in \fig{tau})
caused by the presence of a very small eigenvalue. Although we did
not measure them, this is expected to
produce large fermionic contributions to the driving forces since
they involve an inverse power of the Dirac operator.
%their evaluation requires an inversion of the Dirac operator.

We found the observed \emph{average} $\lambda_{\rm min}$
to be close to its tree--level estimate with Schr\"odinger functional
boundary conditions~\cite{Sint:1993un}.
However, the \emph{smallest} $\lambda_{\rm min}$ is an order of magnitude below that and
we therefore consider these eigenvalues unphysical and will later
establish their nature as cutoff--effects.

Finally, following the procedure of ref.~\cite{DellaMorte:2003jj}, the
absence of global reversibility violations is explicitly verified even
for trajectories resulting in large values of $\Delta H$. Nevertheless
our experience shows that the increased cost of using a smaller
$\delta\tau$ such that no long periods of rejection occur is
more than compensated by the reduction in autocorrelation time of
all observables. The reason is that already a small decrease of the
integration step--size greatly reduces the Hamiltonian violations.
For example, repeating run I with a step--size of $1/20$ instead
of $1/16$, the longest period of rejection was $4$ (instead of
$47$) consecutive trajectories.

\subsection{MC estimates of fermionic observables}

We concluded in the previous section that unphysically small eigenvalues
of $\hat Q^2$ produce algorithmic problems only on a practical
and not on a theoretical level. But apart from slowing down the algorithm
these small eigenvalues also cause problems in the MC evaluation of
fermionic Green's functions. 

Consider the Schr\"odinger functional correlation function $f_1$ as defined
in ref.~\cite{Capitani:1998mq}. It is the correlation between pseudo--scalar
composite fields at the first and last time--slice, respectively. We will
denote its value on a given gauge field configuration by $\ff$.
\begin{figure}[t]
\begin{center}
\includegraphics[angle=0,width=11cm]{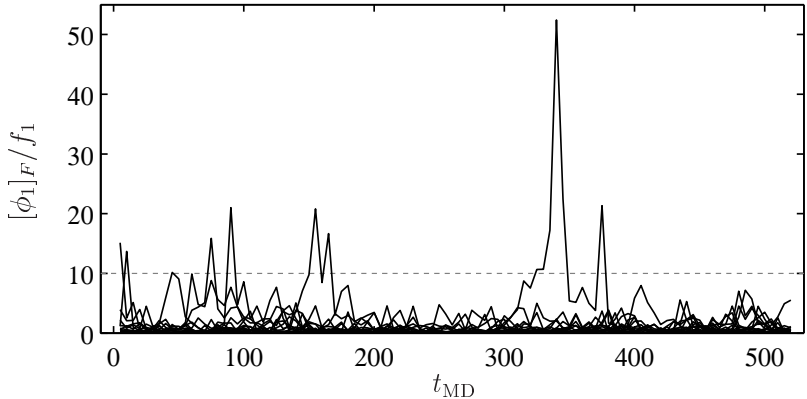}
\end{center}
\vspace{-0.5cm}
\caption{Monte Carlo history for the $N_{\rm rep}\!=\!16$ replica of run II showing
the normalized $\ff$.}
\label{f1}
\end{figure}
\fig{f1} shows the MC history of the normalized $\ff$ for the 16 replica of run II.
Here $t_{\rm MD}$ refers to the molecular dynamics time for each replicum.
While on this scale the bulk of the data are below one and hence not visible
there are several peaks, which have a big influence on the mean value.
These spikes also affect the error estimate $\sigma(f_1)$ through both the variance
and the integrated autocorrelation time \cite{Wolff:2003sm}.
For statistically accessible quantities the error should
approach a $1/\sqrt{t_{\rm MD}}$ behavior in the limit
$t_{\rm MD}\!\rightarrow\!\infty$.
In this respect we found $f_1$ and all other fermionic correlation functions we
considered to be very hard to measure. Even when using $16$ replica,
this asymptotic behavior does not set in after $t_{\rm MD}\!\simeq\!500$. 

\begin{figure}[t]
\begin{center}
\includegraphics[angle=0,width=11cm]{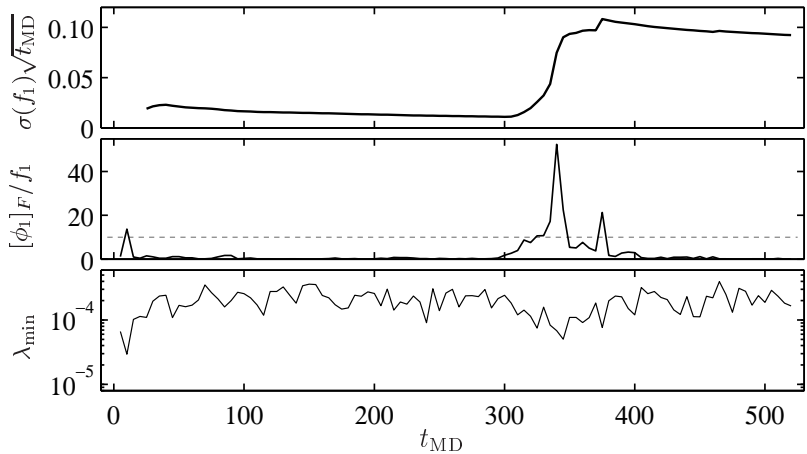}
\end{center}
\vspace{-0.5cm}
\caption{Normalized $\ff$ and  smallest eigenvalue from one ''sick'' replicum
of run II. Evidently the spike in $\ff$ is dominating the statistical error $\sigma(f_1)$.}
\label{f1sick}
\end{figure}

The reason is the rare occurrence of very large values of $\ff$,
which appear to be correlated with small eigenvalues of $\hat Q^2$.
However, this  effect is washed out by using several replica.
We therefore show in \fig{f1sick} the MC history
of $\ff$, $\lambda_{\rm min}$ and our error estimate for $f_1$ 
 for one replicum of run II with such a spike in $\ff$.
Indeed, for each spike in $\ff$ the smallest eigenvalue drops below its
average. That the converse is not true could be ascribed to a lack of overlap 
of the eigenvector corresponding to $\lambda_{\rm min}$ with the source needed
to compute the quark propagator.
Quantitatively, for the correlation between
$\ff$ and $\lambda_{\rm min}$ we measure a value of
$C_{\ff,\lambda_{\rm min}}=-0.33(4)$ if we use all replica and
$-0.46(6)$ from the replicum shown in \fig{f1sick} alone.
Here we used as a definition of the correlation $C_{A,B}$ between two
observables $A$ and $B$

\be
C_{A,B}=\frac{\ev{AB}-\ev A\ev B}
{\sqrt{\Big\langle A^2-\ev A^2\Big\rangle\Big\langle
B^2-\ev B^2\Big\rangle}}\;,\ \ \textrm{ so that }\ \
-1\leq C_{A,B}\leq1\;.
\ee

Even though in the limit of infinite statistics configurations carrying very
small eigenvalues are given the correct weight, depending on the algorithm this
might be badly approximated for a typical ensemble size. 
Similar arguments referring in particular to the HMC algorithm motivated the
introduction of the Polynomial Hybrid Monte Carlo (PHMC) algorithm in
refs.~\cite{Frezzotti:1997ym}.

Hence the difficulty in measuring fermionic correlation functions
might be an efficiency problem related to the choice of the algorithm.
To check this conjecture we employ a second algorithm and compare ensembles
generated by HMC (with two pseudo--fermion fields) with PHMC ensembles.
Indeed, PHMC can be tuned in such a way that it enhances the occurrence of
configurations carrying small eigenvalues, thus resulting in a better sampling
of this region of configuration space. A reweighting step is introduced to
render the algorithm exact. As a preparation for the following discussions
we want to recall some properties and introduce the notations concerning
the PHMC.

\subsubsection{The PHMC algorithm}

One of the main ideas of the PHMC algorithm is to deliberately move away from
importance sampling
by using an approximation to the fermionic part of the lattice QCD action.
More precisely, in an HMC algorithm the inverse of $\hat Q^2$ is replaced
by a polynomial $P_{n,\epsilon}(\hat Q^2)$ of degree $n$.
Here $P_{n,\epsilon}(x)$ approximates
$1/x$ in the range $\epsilon\leq x\leq1$. 
As a consequence this algorithm stochastically implements the weight
$\rmd U\det P^{-1}_{n,\epsilon}(\hat Q^2)e^{-S_g}$,
whereas standard HMC generates ensembles according to $\rmd U\det \hat Q^2e^{-S_g}$
with $S_g$ being the gauge part of the action and $U$ the gauge link configuration.
Denoting averages over the PHMC ensemble by $\langle\dots\rangle_{\rm P}$,
the correct sample average of an observable $\langle O\rangle$ can then be
written as
\be \label{QCD_ave}
\langle { O} \rangle = \langle O\omega \rangle_{\rm P}\;
,\textrm{ where}\ \  \omega=\frac W{\ev W_P}\;,
\ee
and we introduce the reweighting factor $W$ as a (partially)
\footnote{Through the separate treatment of the lowest eigenvalues of $\hat Q^2$
the infrared part of $W$ is evaluated exactly.}
stochastic estimate of $\det\{\hat Q^2 P_{n,\epsilon}(\hat Q^2)\}$.
When using Chebyshev polynomials the relative approximation error for
$\epsilon\leq x\leq1$ is bounded by $\delta \simeq 2 \exp (-2\sqrt{\epsilon}n)$.

\begin{figure}[t]
\begin{center}
\includegraphics[angle=0,width=13cm]{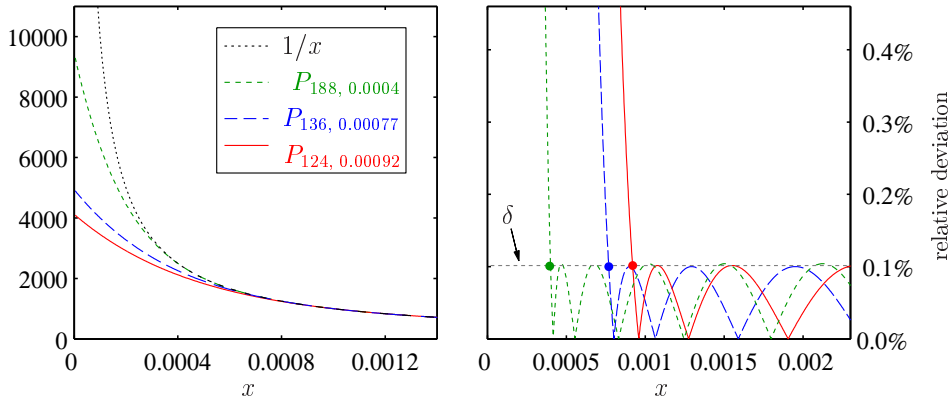}
\end{center}
\vspace{-0.8cm}
\caption{Three different Chebyshev polynomials approximating $1/x$,
all with $\delta=0.001$. The right--hand plot shows the relative
deviation from $1/x$ as a function of $x$. There the points
$(\epsilon,\delta)$ are marked by dots.}
\label{polynomial}
\end{figure}

To give an impression of the r{\^o}le of $\epsilon$ and $\delta$ we plot in
\fig{polynomial} a set of polynomials $P_{n,\epsilon}(x)$ for typical
(in our simulations) values of these parameters and compare them with
$1/x$ in the region of small $x$. Depending on the smallest eigenvalue
of $\hat Q^2$ the parameters $\epsilon$ and $n$ have to be tuned
such that the reweighting factor does not fluctuate too much.
The authors of ref.~\cite{Frezzotti:1997ym} suggested to take $\epsilon$
of the same order as $\langle\lambda_{\rm min}\rangle$ and in practice
used $\epsilon\simeq 2\langle\lambda_{\rm min}\rangle$ and
$\delta\lesssim 0.01$.

Recalling that PHMC replaces  $\det \hat Q^2$ in the  HMC weight with
$\det P^{-1}_{n,\epsilon}(\hat Q^2)$ and observing from \fig{polynomial} that
$P_{n,\epsilon}(x)$ is smaller than $1/x$ for $x\leq\epsilon$,
the aforementioned property of enhancing the occurrence
of small eigenvalues is evident. At this point we would like to
note that the fermionic contribution to the driving force in the PHMC
is bounded from above since $P_{n,\epsilon}(x)$ is finite even at
$x=0$.
In this way the polynomial provides a regularized inversion of
$\hat Q^2$, thus also addressing the problems mentioned in \sect{instab}.

\subsubsection{HMC vs. PHMC}

\begin{figure}[t]
\begin{center}
\includegraphics[angle=0,width=11cm]{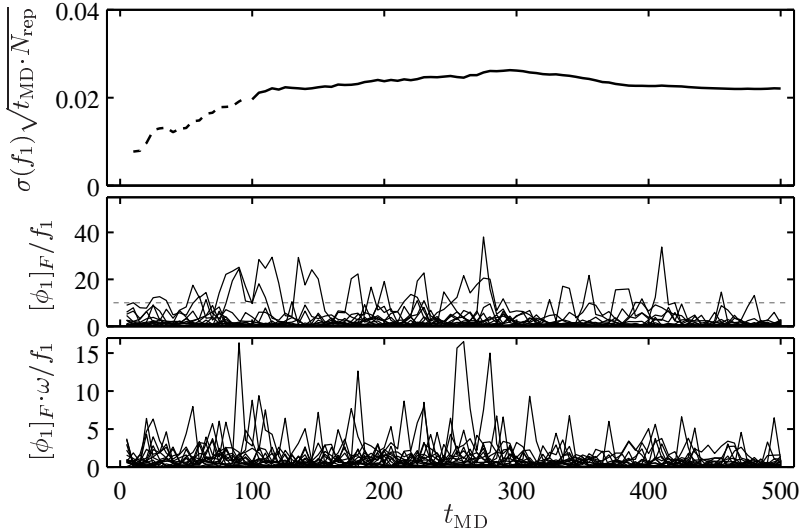}
\end{center}
\vspace{-0.5cm}
\caption{Monte Carlo history for the 16 replica of run III showing
the correlation function $\ff$ and the product $\ff\cdot \omega$,
where $\omega$ is the normalized reweighting factor. Our error
estimate of $f_1$ shows the expected scaling behavior as soon
as the run is long enough for a reliable extraction of $\tau_{\rm int}$.}
\label{reweight}
\end{figure}

Coming back to the comparison of samples from HMC and PHMC, we repeated
run II with PHMC using a polynomial of degree $140$ and
$\epsilon=6\cdot\!10^{-4}$, resulting in $\delta\simeq0.002$. The ratio
$\epsilon/\ev{\lambda_{\rm min}}$ turned out to be around $2.7$.
In \fig{reweight} we plot for this run the MC history of $\ff$ and of
$\ff\cdot \omega$,
which enters into \eq{QCD_ave} if we consider $O=\ff$, i.e.
\be
f_1=\ev \ff=\ev{\ff\cdot \omega}_P=\frac{\ev{\ff\cdot W}_P}{\ev W_P}\;.
\ee
We first observe that apart from 
removing the largest spikes the inclusion of the reweighting factor does not
seem to significantly change the relative fluctuations. This means that
the parameters of the polynomial have been chosen properly.
Events where $\ff$ assumes a value $\rmO(10)$ times larger than
$f_1$ are no longer isolated as in \fig{f1} but happen frequently,
which means that the PHMC algorithm can more easily explore the
associated regions in configuration space.
This is what allows a reliable error estimate as shown in the upper
part of \fig{reweight}, i.e. with 16 replica the asymptotic behavior of
the error sets in after $t_{\rm MD}\!\simeq\!100$.

\begin{figure}[t]
\begin{center}
\includegraphics[angle=0,width=11cm]{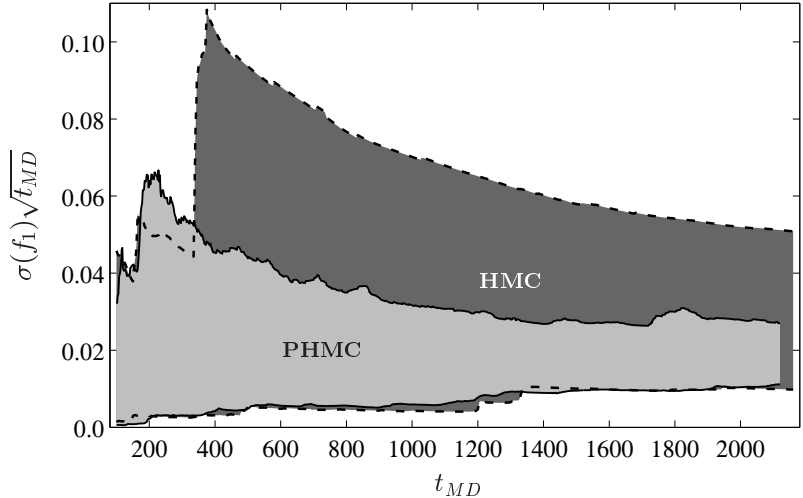}
\end{center}
\vspace{-0.5cm}
\caption{Monte Carlo history of $\sigma(f_1)\sqrt{t_{\rm MD}}$ from
extensions of runs II and III. For the two algorithms we show
the ranges covered by the 16 replica.}
\label{scaling}
\end{figure}

The advantage of using PHMC instead of HMC can be clearly seen by
considering the spread of $\sigma(f_{1})\sqrt{t_{\rm MD}}$ among
the replica. We analyzed this quantity in extensions of runs II
and III. The result is shown in figure \fig{scaling}, where the
shaded areas represent the range of values covered by the 16 replica
as a function of the MD time.
In the limit of infinite statistics all replica should converge to the
same value, which need not be the same for the two algorithms because of
reweighting and different autocorrelation times. We see
that the spread for the HMC data is more than twice as large as for
PHMC, i.e. the error on $f_1$ is significantly harder to estimate with
HMC.
What we are suggesting here is that the algorithm should be chosen
depending on the type of observables and the parameter values.
From our experience we conclude
that PHMC sampling might just be more effective than HMC when computing
fermionic 
quantities that are sensitive to small eigenvalues. 

To gain some more insight into the difference in sampling we consider the
distribution of $\lambda_{\rm min}$ since this is where we expect the largest
effect. The distributions are analyzed by treating
$\Lambda_{\rm bin}\!=\!\chi_{\rm bin}(\lambda_{\rm min})$ as a primary observable. Here
$\chi_{\rm bin}$ denotes the characteristic function of each given bin in the
histogram.
We then perform our normal error analysis for
$\ev{\Lambda_{\rm bin}}$,
where \eq{QCD_ave} has to be used if it is a
PHMC sample. For comparison $\ev{\Lambda_{\rm bin}}_{\rm P}$ is also analyzed in
this case.

\begin{figure}[t]
\begin{center}
\includegraphics[angle=0,width=12cm]{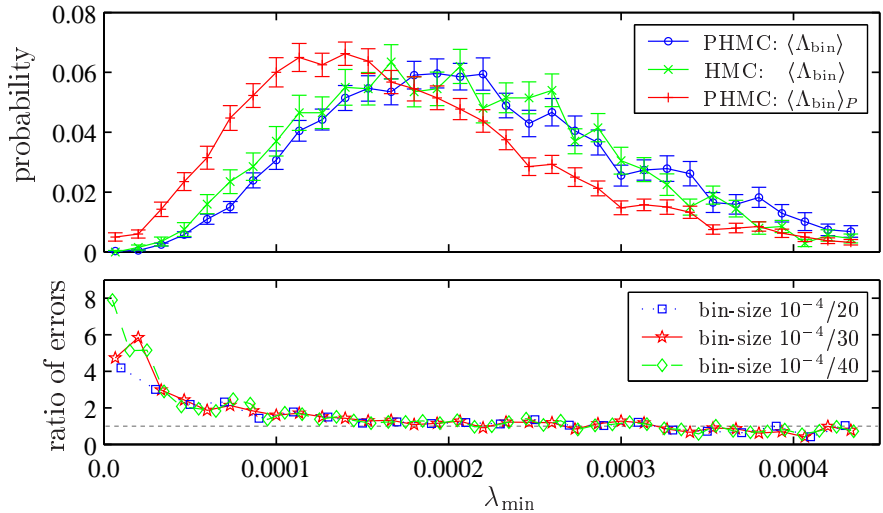}
\end{center}
\vspace{-0.5cm}
\caption{Upper panel: histograms of $\lambda_{\rm min}$, i.e. 
$\ev{\Lambda_{\rm bin}}$ vs. 'bin', from runs II and III. For run
III we also show $\ev{\Lambda_{\rm bin}}_P$.
Lower panel (from the same data): ratio of the error on $\ev{\Lambda_{\rm bin}}$
from HMC to that from PHMC.}
\label{compsample}
\end{figure}

The histograms in the upper part of \fig{compsample} compare the results
from 200 independent measurements produced by HMC and PHMC (runs II and III,
respectively). As expected the distributions agree within errors.
For the PHMC run we also plot the unreweighted histogram,
i.e. $\ev{\Lambda_{\rm bin}}_{\rm P}$.
Here we again confirm that with the parameters we chose for the polynomial
the PHMC produces more configurations with small eigenvalues than HMC.
As a consequence
of the reweighting the errors at the infrared end of the spectrum should be
smaller for the PHMC data. This is explicitly verified in the lower part
of the plot where we show the ratio of the errors on $\ev{\Lambda_{\rm bin}}$
from the two algorithms. The three symbols refer to different bin sizes. 
The advantage in using PHMC to sample this part of the spectrum is
significant and we will make use of this in the following discussion.

\section{Comparison to the quenched case}\label{sect_three}

In the previous section we studied various problems related
to the occurrence of small eigenvalues. All the data presented
there were produced at bare parameter values, which correspond
to relatively large quark masses and small volumes.
These small eigenvalues might therefore have a different nature from
the ''physical'' ones expected to show up in large volumes and/or
close to the chiral limit. Here and in the next section we will establish
them as cutoff--effects.

\begin{figure}[t]
\begin{center}
\includegraphics[angle=0,width=12cm]{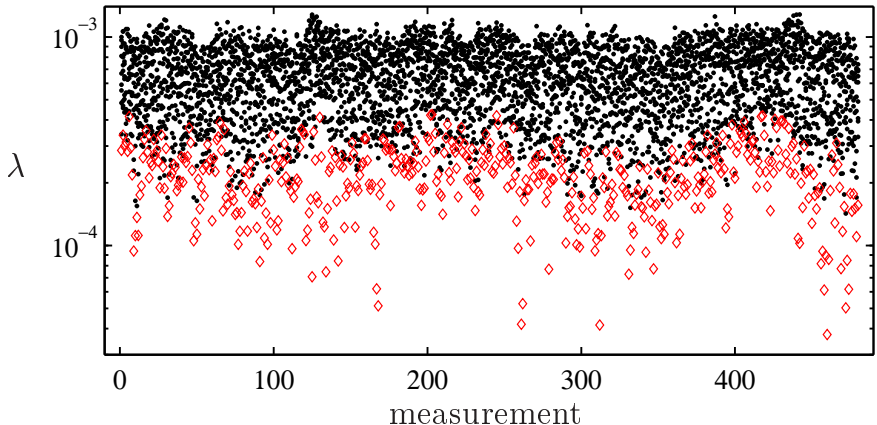}
\end{center}
\vspace{-0.5cm}
\caption{Monte Carlo history of the ten lowest eigenvalues 
at the parameters of run II. The open symbols denote $\lambda_{\rm min}$.
} 
\label{bulk}
\end{figure}

To this end we made an additional simulation at the parameters of run II and
calculated the ten lowest--lying eigenvalues
$\lambda_i$, $i=1\ldots10$.
In \fig{bulk} the smallest eigenvalue, $\lambda_1$, is denoted by an open
symbol. It seems that while $\lambda_2$ through $\lambda_{10}$ form a
rather compact band, the lowest eigenvalue fluctuates to very small
values quite independently of the others. It is expected and has been
shown numerically \cite{Hernandez:1998et} that the spectrum of the
Dirac operator depends quite strongly on the bare gauge coupling.
A well--defined lower bound should be recovered close to the continuum
limit only. Therefore we take the strong fluctuations of $\lambda_{\rm min}$
as an indication for the presence of large cutoff--effects. Here we should
point out that the eigenvalues of the Dirac operator are not on--shell 
quantities. Hence the Symanzik improvement programme does not necessarily
reduce cutoff--effects here. Quenched experience even suggests 
that the opposite might be true \cite{DeGrand:1998mn}.

The occurrence of small eigenvalues at these bare parameters
poses a somewhat unexpected problem in dynamical simulations.
Comparing the quenched situation to the $N_f\!=\!2$ dynamical case, the 
na\"{\i}ve expectation is that at fixed bare parameters the
probability of finding configurations with small eigenvalues should be reduced
by the determinant.
%This will result in a shift of  $\ev {\lambda_{\rm min}}$ to higher
%values. Consider two gauge configurations
%$U_1$ and $U_2$. Depending on the number of dynamical quarks $N_f$
%their relative weight in the path integral is given by
%
%\be
%R_{12}(N_f)=\frac{e^{-S_g[U_1]}}{e^{-S_g[U_2]}}\left(\frac{\det Q[U_1]}
%{\det Q[U_2]}\right)^{N_f}\; ,
%\ee
%so that
%\be
%\frac{R_{12}(N_f\!=\!2)}{R_{12}(N_f\!=\!0)}=\frac{\det Q^2[U_1]}{\det Q^2[U_2]}=
%\prod_i\frac{\lambda_i[U_1]}{\lambda_i[U_2]}\; .
%\ee
%
%With the assumption that significant fluctuations of the determinant
%come from the small eigenvalues only, one can conclude that at fixed
%bare parameters the suppression
%of configurations with a small eigenvalue should be proportional to the
%latter.
To us the more relevant question seems to be whether small eigenvalues are suppressed
in a situation where the physical parameters (e.g. volume and
pseudo--scalar mass) are kept constant.

\begin{figure}[t]
\begin{center}
\includegraphics[angle=0,width=12cm]{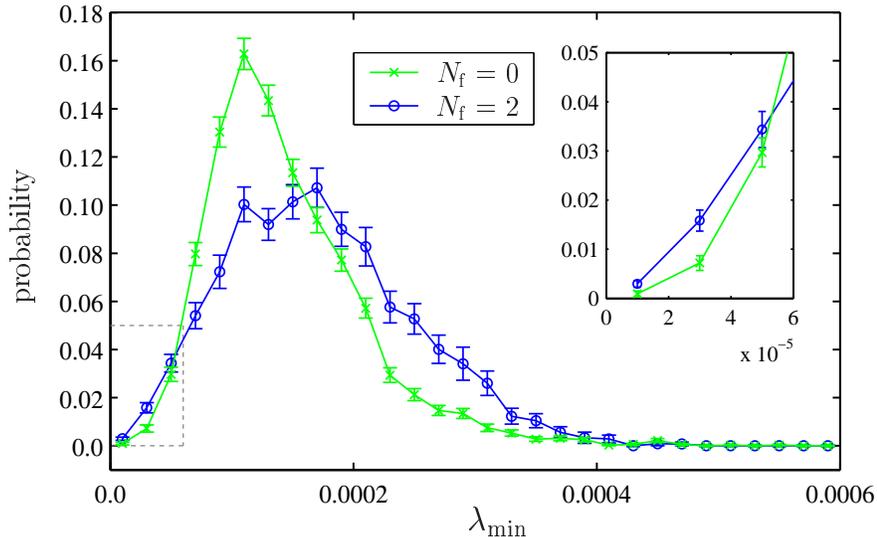}
\end{center}
\vspace{-0.5cm}
\caption{The histograms of $\lambda_{\rm min}$ from run IV ($N_f\!=\!2$) and
run V ($N_f\!=\!0$). Despite its higher mean value the dynamical data show
a much larger probability of finding very small eigenvalues.}
\label{cutoff}
\end{figure}

Using the quenched data from ref.~\cite{Garden:1999fg} and the dynamical data
from refs.~\cite{Aoki:2002uc} and \cite{Allton:2001sk}
(where an estimate of $r_0/a\!=\!5.21(6)$ for $\beta\!=\!5.2$ can be found)
we chose the parameters of the quenched run V
such that the lattice spacing and the (large volume) pseudo--scalar mass
are matched to run IV.
This was found to occur at almost equal bare current quark mass
(see $Lm$ in \tab{t_simpar}).
In \fig{cutoff} we compare the distributions of
$\lambda_{\rm min}$ for these two runs. Two comments are in order here:
\begin{itemize}
\item
For the dynamical run the mean value is shifted
up from $1.44(1)\!\cdot\!10^{-4}$ to $1.72(5)\!\cdot\!10^{-4}$. 
This agrees with the na\"{\i}ve expectation but in a physically matched 
comparison it is a non--trivial observation.
\item
The distribution itself is
significantly broader compared to the quenched case and in particular it is 
falling off more slowly towards zero. This means that even though
$\ev{\lambda_{\rm min}}$ is larger for $N_f\!=\!2$ the probability of finding
\emph{very} small eigenvalues is enhanced.
\end{itemize}

The second point, i.e. that the lower bound of $\lambda_{\rm min}$ is 
less well--defined, seems to imply that at a
lattice spacing of $a\approx0.1\fm$ the cutoff--effects are much larger
in the $N_f\!=\!2$ case. 
To substantiate this we will compare the
distribution of $\lambda_{\rm min}$ to that from a run at finer lattice spacing
and matched physical parameters.

\section{Finer lattices}\label{sect_four}

Apart from cutoff--effects, in the massless theory the Schr\"odinger functional
coupling $\bar g^2$ is a
function of the box size $L$ only \cite{Luscher:1992an,Sint:1993un}.
We measured it on a small lattice of extension $L/a\!=\!4$ at $\beta=5.2$,
obtaining a value of $\bar g^2\!=\!3.7(1)$. We then extrapolated to this value the
$L/a\!=\!6$ data used in ref.~\cite{Bode:2001jv} as a function of $\beta$.
Our result from the matching
is that for the two--flavor theory a bare gauge coupling of $\beta\!=\!5.5$
roughly corresponds to a lattice spacing, which is 1.5 times smaller than at
$\beta\!=\!5.2$.

Hoping that the algorithmic difficulties arising from cutoff--effects
would be much smaller in this situation, we simulated a $12^3\!\times\!27$
lattice at this value of $\beta$ (run VI) using the HMC algorithm.
With the $\kappa $ we chose (and ignoring the change in
renormalization factors) the bare quark mass $Lm$ is roughly
matched to the heavier runs at $\beta\!=\!5.2$.
We therefore compare run VI with run III.

Normally, a constant acceptance requires a decrease of the MD
integration step--size if ones goes to finer lattices at fixed
physical conditions. This argument is based on the scaling of
the small eigenvalues, which influence the MD driving force.
We found that $\ev{\lambda_{\rm min}}$ in run VI is a factor two
smaller than in run III.
Nevertheless, at $\beta\!=\!5.5$ the step--size necessary for
a certain ($\simeq90\%$) acceptance is roughly the same as at $\beta\!=\!5.2$.
This indicates that the value of $\delta\tau$ we had to use in the HMC runs
at $\beta\!=\!5.2$ was dictated by the occurrence of extremely small
eigenvalues rather than by the average smallest eigenvalue.
In addition, where in run I at the same average
acceptance a maximum of $47$ proposals were rejected in a row, the maximum
for run VI is $4$ trajectories. For this reason $e^{-\Delta H}$ shows no
autocorrelation and its distribution is well separated from zero.

Concerning fermionic observables, we have not observed spikes and hence
expect the error to scale properly. However, for an accurate estimate of
the error on e.g. $f_1$ our present statistics is not yet
sufficient.\footnote{Ratios of correlators relevant for physical applications are
easier to estimate.}

%However, due to extremely long
%autocorrelation times ($\gtrsim50$ trajectories) the statistics we have
%accumulated in this exploratory run is not
%sufficient to provide a reliable error on e.g. $f_1$.

\begin{figure}[t]
\begin{center}
\includegraphics[angle=0,width=12cm]{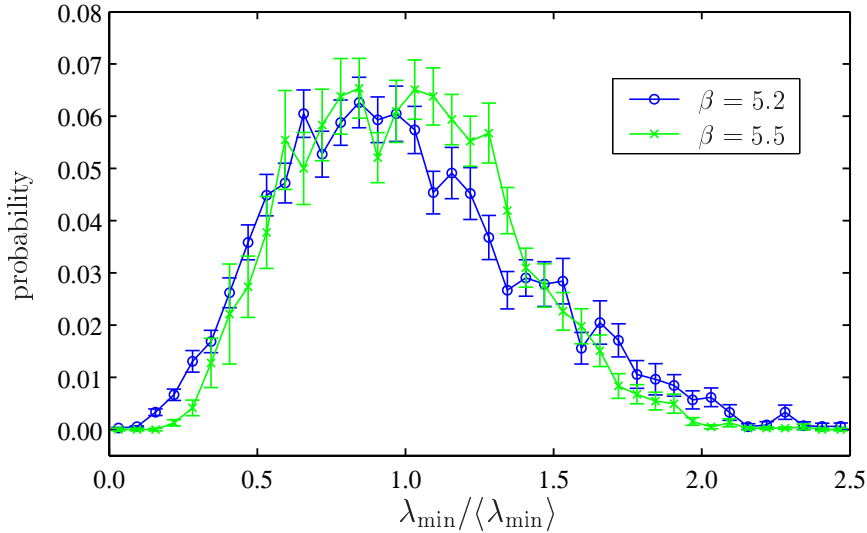}
\end{center}
\vspace{-0.5cm}
\caption{Normalized distributions of $\lambda_{\rm min}$ from runs III
($\beta\!=\!5.2$) and VI ($\beta\!=\!5.5$). While the data from the coarse lattice
stretch almost to zero, the $\beta\!=\!5.5$ data seem to have a more well--defined
lower bound.}
\label{roughmatch}
\end{figure}

The reason for these effects is the change in the distribution of
$\lambda_{\rm min}$. To compensate for the different lattice spacing,
\fig{roughmatch} compares $\lambda_{\rm min}/\ev{\lambda_{\rm min}}$
from runs III and VI. One can clearly see that at the finer lattice spacing
the probability of finding a smallest eigenvalue less than half its average
is greatly reduced compared to $\beta\!=\!5.2$. The width of the distribution
is smaller in this case and in particular the spectrum is now clearly
separated from zero. Quantitatively, the normalized variance of
$\lambda_{\rm min}$ is reduced from $0.18(1)$ to $0.13(2)$.

\begin{figure}[t]
\begin{center}
\includegraphics[angle=0,width=12cm]{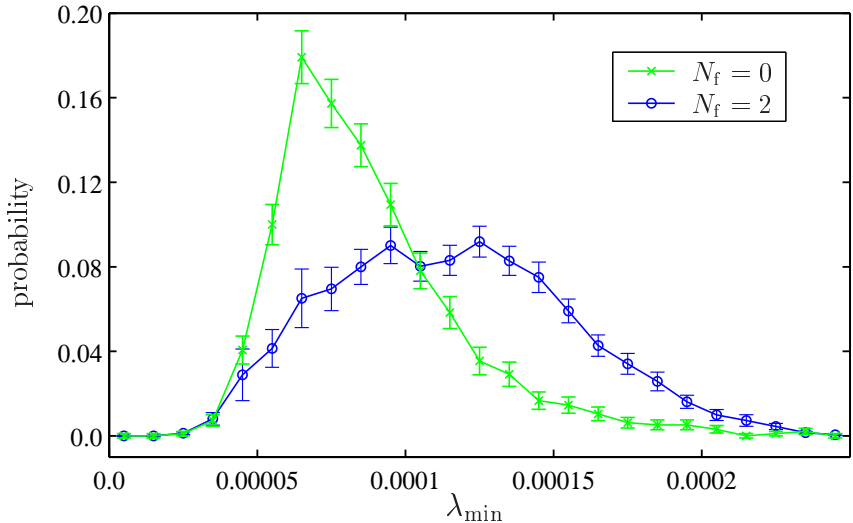}
\end{center}
\vspace{-0.5cm}
\caption{The histograms of $\lambda_{\rm min}$ from run VI ($N_f\!=\!2$) and
run VII ($N_f\!=\!0$). At this finer lattice spacing the
lower end of the spectrum
appears to be similar in the quenched and the dynamical case.
}
\label{matchfine}
\end{figure}

This comparison explicitly shows that the long tail of the eigenvalue
distribution we observed at $a\simeq0.1\fm$, and which caused the
problems we have discussed, is a cutoff-effect. Matching also run VI
to a quenched simulation (run VII), we again found an upward shift of
$\ev{\lambda_{\rm min}}$ for the dynamical case. In addition, at this finer
lattice spacing, the tails of the distributions of $\lambda_{\rm min}$ look
already very similar to each other as shown in \fig{matchfine}.

\section{Conclusions}

At a lattice cutoff of approximately $2\;\rm{GeV}$ we have
studied the behavior and performance of HMC--type algorithms
in an intermediate size volume of $1\fm^4$.
We discussed problems related to the occurrence of
small eigenvalues in two--flavor dynamical simulations with improved
Wilson fermions. We found these small eigenvalues to be responsible
for large Hamiltonian violations in the molecular dynamics. Even for
integration step--sizes such that the acceptance is $80\sim90\%$ those
can still cause long periods of rejection, thus degrading algorithmic
performance. However, in spite of employing only single--precision
arithmetics we never observed
reversibility violations.

In addition, those eigenvalues make the estimate of fermionic
quantities very difficult. The na\"{\i}ve intuition is that the fermionic determinant
should suppress small eigenvalues compared to the quenched case.
Through a direct comparison at matched physical parameters we indeed verified
that $\ev{\lambda_{\rm min}}$ is larger with two dynamical flavors. On the other
hand there is no obvious expectation for the tail of the distribution
and we observed that it extends further towards zero than in the quenched case.
Given the infrared cutoff induced by the Schr\"odinger functional boundary
conditions and the quark mass
we interpret this as a lattice artifact. We were able to confirm this
picture with a simulation at finer lattice spacing, where the spectrum
turned out to have a much sharper lower bound.

In our study we found that the PHMC algorithm is more efficient than
HMC (with two pseudo--fermions) in incorporating the contribution to
the path integral of configurations
carrying small eigenvalues. In other words, the distortion of the spectrum by
cutoff--effects actually makes it advantageous to deviate from
importance sampling.
Also without such special problems we found PHMC at least comparable
in performance to HMC (in our implementations).

We want to emphasize that the problems discussed here do not
occur only in the Schr\"odinger functional setup. Without
this infrared regulator they are expected to show
up already at larger quark masses.\\[0.5cm]

%These findings seem to support the suggested proximity to a phase
%transition at the parameter value we explored with the coarse lattices.
%A scaling study will enable us to explore this in
%a more quantitative way and also help to judge the benefits of
%new action with respect to the problems reported.\\[0.5cm]

{\bf Acknowledgements.} We are grateful to R.~Sommer for discussions and
a critical reading of the manuscript.
Discussions with S.~Aoki, S.~Hashimoto and T.~Kaneko are also acknowledged.
We thank NIC/DESY Zeuthen for allocating computer time on the APEmille
machines indispensable to this project and the APE group for their
professional and constant support.
This work is supported in part by the Deutsche Forschungsgemeinschaft
in the SFB/TR 09-03, ``Computational Particle Physics'' and the
Graduiertenkolleg GK271.


\begin{thebibliography}{99}

%\cite{Aoki:2001xq}
\bibitem{Aoki:2001xq}
S.~Aoki {\it et al.}  [JLQCD Collaboration],
%``Non-trivial phase structure of N(f) = 3 QCD with O(a)-improved Wilson
%fermion at zero temperature,''
Nucl.\ Phys.\ Proc.\ Suppl.\  {\bf 106} (2002) 263.
%%CITATION = HEP-LAT 0110088;%%

%\cite{Sommer:2003ne}
\bibitem{Sommer:2003ne}
R.~Sommer {\it et al.}  [ALPHA Collaboration],
%``Large cutoff effects of dynamical Wilson fermions,''
Nucl.\ Phys.\ Proc.\ Suppl.\  {\bf 129-130} (2004) 405.
%%CITATION = HEP-LAT 0309171;%%

%\cite{Heitger:1999dw}
\bibitem{Heitger:1999dw}
J.~Heitger  [ALPHA Collaboration],
%``Scaling investigation of renormalized correlation functions in O(a)  improved
%quenched lattice QCD,''
Nucl.\ Phys.\ B {\bf 557} (1999) 309.
%%CITATION = HEP-LAT 9903016;%%

%\cite{Hoffmann:2003mm}
\bibitem{Hoffmann:2003mm}
R.~Hoffmann {\it et al.} [ALPHA Collaboration],
%``Non-perturbative renormalization of the axial current with improved Wilson
%quarks,''
Nucl.\ Phys.\ Proc.\ Suppl.\  {\bf 129-130} (2004) 423.
%%CITATION = HEP-LAT 0309071;%%

%\cite{Durr:2003nc}
\bibitem{Durr:2003nc}
S.~D\"urr and M.~Della Morte,
%``Exploring two non-perturbative definitions of c(A),''
Nucl.\ Phys.\ Proc.\ Suppl.\  {\bf 129-130} (2004) 417.
%%CITATION = HEP-LAT 0309169;%%

%\cite{deForcrand:1996ck}
\bibitem{deForcrand:1996ck}
P.~de Forcrand and T.~Takaishi,
%``Fast fermion Monte Carlo,''
Nucl.\ Phys.\ Proc.\ Suppl.\  {\bf 53} (1997) 968.
%%CITATION = HEP-LAT 9608093;%%

%\cite{Frezzotti:1997ym}
\bibitem{Frezzotti:1997ym}
R.~Frezzotti and K.~Jansen,
%``A polynomial hybrid Monte Carlo algorithm,''
Phys.\ Lett.\ B {\bf 402} (1997) 328.\\
%%CITATION = HEP-LAT 9702016;%%
R.~Frezzotti and K.~Jansen,
%``The PHMC algorithm for simulations of dynamical fermions.  I: Description
%and properties,''
Nucl.\ Phys.\ B {\bf 555} (1999) 395.\\
%%CITATION = HEP-LAT 9808011;%%
R.~Frezzotti and K.~Jansen,
%``The PHMC algorithm for simulations of dynamical fermions. II:  Performance
%analysis,''
Nucl.\ Phys.\ B {\bf 555} (1999) 432.
%%CITATION = HEP-LAT 9808038;%%


\bibitem{Luscher:1992an}
M.~L\"uscher, R.~Narayanan, P.~Weisz and U.~Wolff,
%``The Schrodinger functional: A Renormalizable probe for nonAbelian gauge
%theories,''
Nucl.\ Phys.\ B {\bf 384} (1992) 168.
%%CITATION = HEP-LAT 9207009;%%

%\cite{Sint:1993un}
\bibitem{Sint:1993un}
S.~Sint,
%``On the Schrodinger functional in QCD,''
Nucl.\ Phys.\ B {\bf 421} (1994) 135.
%%CITATION = HEP-LAT 9312079;%%

%\cite{Luscher:1996sc}
\bibitem{Luscher:1996sc}
M.~L\"uscher, S.~Sint, R.~Sommer and P.~Weisz,
%``Chiral symmetry and O(a) improvement in lattice QCD,''
Nucl.\ Phys.\ B {\bf 478} (1996) 365.
%%CITATION = HEP-LAT 9605038;%%

%\cite{Luscher:1996ug}
\bibitem{Luscher:1996ug}
M.~L\"uscher {\it et al.} [ALPHA Collaboration],
%``Non-perturbative O(a) improvement of lattice QCD,''
Nucl.\ Phys.\ B {\bf 491} (1997) 323.
%%CITATION = HEP-LAT 9609035;%%

%\cite{Jansen:1998mx}
\bibitem{Jansen:1998mx}
K.~Jansen and R.~Sommer  [ALPHA Collaboration],
%``O(alpha) improvement of lattice QCD with two flavors of Wilson quarks,''
Nucl.\ Phys.\ B {\bf 530} (1998) 185
[Erratum-ibid.\ B {\bf 643} (2002) 517].
%%CITATION = HEP-LAT 9803017;%%

%\cite{Yamada:2004ja}
\bibitem{Yamada:2004ja}
N.~Yamada {\it et al.}  [JLQCD Collaboration],
%``Non-perturbative O(a)-improvement of Wilson quark action in three-flavor QCD
%with plaquette gauge action,''
hep-lat/0406028.
%%CITATION = HEP-LAT 0406028;%%

%\cite{Bode:1999sm}
\bibitem{Bode:1999sm}
A.~Bode, P.~Weisz and U.~Wolff  [ALPHA Collaboration],
%``Two loop computation of the Schroedinger functional in lattice QCD,''
Nucl.\ Phys.\ B {\bf 576} (2000) 517
[Erratum-ibid.\ B {\bf 600} (2001) 453 and B {\bf 608} (2001) 481].
%%CITATION = HEP-LAT 9911018;%%

%\cite{Sint:1997jx}
\bibitem{Sint:1997jx}
S.~Sint and P.~Weisz,
%``Further results on O(a) improved lattice QCD to one-loop order of
%perturbation theory,''
Nucl.\ Phys.\ B {\bf 502} (1997) 251.
%%CITATION = HEP-LAT 9704001;%%

%\cite{Luscher:1996vw}
\bibitem{Luscher:1996vw}
M.~L\"uscher and P.~Weisz,
%``O(a) improvement of the axial current in lattice QCD to one-loop order  of
%perturbation theory,''
Nucl.\ Phys.\ B {\bf 479} (1996) 429.
%%CITATION = HEP-LAT 9606016;%%

%\cite{Hasenbusch:2001ne}
\bibitem{Hasenbusch:2001ne}
M.~Hasenbusch,
%``Speeding up the Hybrid-Monte-Carlo algorithm for dynamical fermions,''
Phys.\ Lett.\ B {\bf 519} (2001) 177.\\
%%CITATION = HEP-LAT 0107019;%%
M.~Hasenbusch and K.~Jansen,
%``Speeding up lattice QCD simulations with clover-improved Wilson fermions,''
Nucl.\ Phys.\ B {\bf 659} (2003) 299.
%%CITATION = HEP-LAT 0211042;%%

\bibitem{DellaMorte:2003jj}
M.~Della Morte  {\it et al.} [ALPHA Collaboration],
%``Simulating the Schroedinger functional with two pseudo-fermions,''
Comput.\ Phys.\ Commun. {\bf 156} (2003) 62.
%%CITATION = HEP-LAT 0307008;%%

\bibitem{Capitani:1998mq}
S.~Capitani, M.~L\"uscher, R.~Sommer and H.~Wittig  [ALPHA Collaboration],
%``Non-perturbative quark mass renormalization in quenched lattice QCD,''
Nucl.\ Phys.\ B {\bf 544} (1999) 669.
%%CITATION = HEP-LAT 9810063;%%

\bibitem{Joo:2000dh}
B.~Joo {\it et al.} [UKQCD Collaboration],
%``Instability in the molecular dynamics step of hybrid Monte Carlo in
%dynamical fermion lattice QCD simulations,''
Phys.\ Rev.\ D {\bf 62} (2000) 114501.
%%CITATION = HEP-LAT 0005023;%%

\bibitem{Wolff:2003sm}
U.~Wolff  [ALPHA Collaboration],
%``Monte Carlo errors with less errors,''
Comput.\ Phys.\ Commun.\  {\bf 156} (2004) 143.
%%CITATION = HEP-LAT 0306017;%%

\bibitem{Namekawa:2002pv}
Y.~Namekawa {\it et al.}  [CP-PACS Collaboration],
%``Exploring QCD at small sea quark masses with improved Wilson-type quarks,''
Nucl.\ Phys.\ Proc.\ Suppl.\  {\bf 119} (2003) 335.\\
%%CITATION = HEP-LAT 0209073;%%
Y.~Namekawa {\it et al.}  [CP-PACS Collaboration],
%``Light hadron spectroscopy in two-flavor QCD with small sea quark masses,''
hep-lat/0404014.
%%CITATION = HEP-LAT 0404014;%%

\bibitem{Allton:2004qq}
C.~R.~Allton {\it et al.}  [UKQCD Collaboration],
%``Improved Wilson QCD simulations with light quark masses,''
hep-lat/0403007.
%%CITATION = HEP-LAT 0403007;%%

%\cite{Frezzotti:2002iv}
\bibitem{Frezzotti:2002iv}
R.~Frezzotti,
%``Wilson fermions with chirally twisted mass,''
Nucl.\ Phys.\ Proc.\ Suppl.\  {\bf 119} (2003) 140.
%%CITATION = HEP-LAT 0210007;%%

%\cite{Hernandez:1998et}
\bibitem{Hernandez:1998et}
P.~Hernandez, K.~Jansen and M.~L\"uscher,
%``Locality properties of Neuberger's lattice Dirac operator,''
Nucl.\ Phys.\ B {\bf 552} (1999) 363.
%%CITATION = HEP-LAT 9808010;%%

%\cite{DeGrand:1998mn}
\bibitem{DeGrand:1998mn}
T.~DeGrand, A.~Hasenfratz and T.~G.~Kovacs,
%``Instantons and exceptional configurations with the clover action,''
Nucl.\ Phys.\ B {\bf 547} (1999) 259.
%%CITATION = HEP-LAT 9810061;%%

%\cite{Garden:1999fg}
\bibitem{Garden:1999fg}
J.~Garden, J.~Heitger, R.~Sommer and H.~Wittig  [ALPHA Collaboration],
%``Precision computation of the strange quark's mass in quenched QCD,''
Nucl.\ Phys.\ B {\bf 571} (2000) 237.
%%CITATION = HEP-LAT 9906013;%%

%\cite{Aoki:2002uc}
\bibitem{Aoki:2002uc}
S.~Aoki {\it et al.}  [JLQCD Collaboration],
%``Light hadron spectroscopy with two flavors of O(a)-improved dynamical
%quarks,''
Phys.\ Rev.\ D {\bf 68} (2003) 054502.
%%CITATION = HEP-LAT 0212039;%%

\bibitem{Allton:2001sk}
C.~R.~Allton {\it et al.}  [UKQCD Collaboration],
%``Effects of non-perturbatively improved dynamical fermions in QCD at  fixed
%lattice spacing,''
Phys.\ Rev.\ D {\bf 65}, 054502 (2002).
%%CITATION = HEP-LAT 0107021;%%

%\cite{Bode:2001jv}
\bibitem{Bode:2001jv}
A.~Bode {\it et al.}  [ALPHA Collaboration],
%``First results on the running coupling in QCD with two massless flavors,''
Phys.\ Lett.\ B {\bf 515}, 49 (2001).
%%CITATION = HEP-LAT 0105003;%%

M.~Della Morte {\it et al.} [ALPHA Collaboration],
%``Recent results on the running coupling in QCD with two massless flavours,''
Nucl.\ Phys.\ Proc.\ Suppl.\  {\bf 119} (2003) 439.
%%CITATION = HEP-LAT 0209023;%%

\end{thebibliography}
\end{document}